\documentclass{pnastwo}
\usepackage{ulem}

\usepackage{amsmath}

\url{www.pnas.org/cgi/doi/10.1073/pnas.0709640104}
\copyrightyear{2008}
\issuedate{Issue Date}
\volume{Volume}
\issuenumber{Issue Number}

\begin{document}

\title{Wavy membranes and the growth rate of a planar chemical garden: Enhanced diffusion and bioenergetics
}

\author{Yang Ding\affil{1}{Department of Chemical Engineering and Biotechnology, University of Cambridge,  Cambridge, UK},
Bruno Batista\affil{2}{Department of Chemistry and Biochemistry, Florida State University, Tallahassee, Florida 32306-4390, USA},
Oliver Steinbock\affil{2}{}, 
Julyan H. E. Cartwright\affil{3}{Instituto Andaluz de Ciencias de la Tierra, CSIC--Universidad de Granada, E-18100 Armilla, Granada, Spain}\affil{4}{Instituto Carlos I de F\'{\i}sica Te\'orica y Computacional, Universidad de Granada, E-18071 Granada, Spain},
\and Silvana S. S. Cardoso\affil{1}{}
}

\contributor{Submitted to Proceedings of the National Academy of Sciences
of the United States of America}

\significancetext{In hydrothermal vents on the ocean floor, precipitation membranes grow at the boundary between seawater and mineral-rich liquid flowing out of the vent. Such membranes are increasingly viewed as having played a vital role in the emergence of life on Earth, but their bioenergetics is unclear. Here, we present a laboratory and theoretical study that quantifies ionic transport across an analog membrane. We demonstrate that flow over a growing, wavy-membrane topography enhances diffusive transport across its surface. This enhanced diffusion helps to explain the ``leakiness'' present in early protocells from chemical gardens. More generally, the work is of interest in fluid-flow control via surface topography, and the opposite: predesigned flow perturbations to shape membrane formation, in biology, chemistry, and physics.
}

\maketitle

\begin{article}

\begin{abstract}
In order to model ion transport across protocell membranes in Hadean hydrothermal vents, we consider both theoretically and experimentally the planar growth of a precipitate membrane formed at the interface between two parallel fluid streams in a two-dimensional microfluidic reactor. The growth rate of the precipitate is found to be proportional to the square root of time, which is characteristic of diffusive transport. However, the dependence of the growth rate on the concentrations of hydroxide and metal ions is approximately linear and quadratic, respectively. We show that such a difference in ionic transport dynamics arises from the enhanced transport of metal ions across a thin gel layer present at the surface of the precipitate. The fluctuations in transverse velocity in this wavy porous gel layer allow an enhanced transport of the cation, so that the effective diffusivity is about an order of magnitude higher than that expected from molecular diffusion alone. Our theoretical predictions are in excellent agreement with our laboratory measurements of the growth of a manganese hydroxide membrane in a microfluidic channel, and this enhanced transport is thought to have been needed to account for the bioenergetics of the first single-celled organisms.
\end{abstract}

\keywords{chemical gardens | origin of life | hydrothermal vents | chemobrionics}

\begin{figure}[tb]
\begin{center}
\includegraphics[width = 0.8\columnwidth]{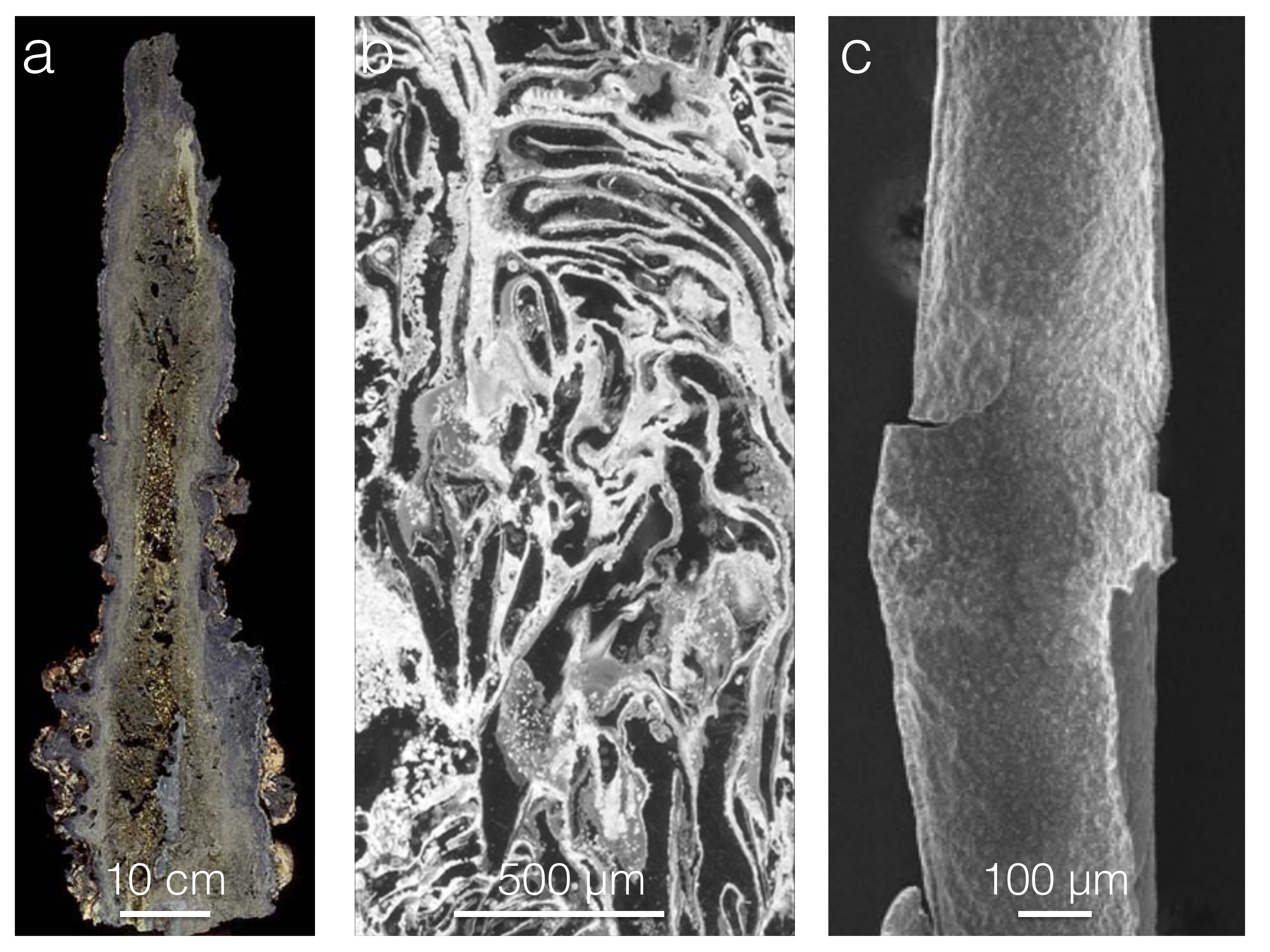}
\end{center}
\caption{Wavy membranes in (a,b) hydrothermal vents and (c) a chemical garden. (a) Longitudinal section of the Olivia chimney from the East Pacific Rise. Metals such as copper, zinc and iron dissolved in the expelled hot fluids precipitate to form the metallic centre of the chimney of this acidic black smoker [13] (Photograph provided by C.E.J. de Ronde, GNS Science). (b) A section through an alkaline white smoker hydrothermal vent wall of Lost City \cite{kelley2005}, showing its microporous structure (From Kelley et al.\cite{kelley2005}), reprinted with permission from AAAS.). (c) Scanning electron micrograph of the wavy, interior surface of a silica-zinc-(hydr)oxide tube formed by injection of a zinc sulfate solution into a host solution of sodium silicate.
\label{fig:wavy}}
\end{figure}

\dropcap{I}N hydrothermal vents, precipitation membranes grow at the boundary between seawater and the mineral-rich liquid flowing out of the vent. The same sort of semipermeable precipitation membrane forms in a laboratory setting in chemical gardens, as well as in other chemobrionic systems \cite{barge2015} (Fig.~\ref{fig:wavy}). Such membranes are increasingly viewed as having played a vital role in the emergence of life on Earth over 4 billion years ago \cite{russell2008,russell2014}. Both acidic `black smoker' (Fig.~\ref{fig:wavy}a) and alkaline `white smoker' (Fig.~\ref{fig:wavy}b) vents display a complex internal structure of precipitation membranes. It is from the geochemistry of the alkaline white smoker vents such as those of the Lost City that, it is thought, the biochemistry of life may most plausibly have arisen \cite{martin2007, sojo2016} and much experimental work is being carried out along these lines \cite{barge2014, herschy2014}.
It has been suggested that LUCA, the Last Universal Common Ancestor, i.e., the progenitor of all present life on Earth, needed to possess a `leaky membrane'  to function in bioenergetic terms in the hydrothermal vent environment \cite{lane2014}. 
Here we demonstrate that flow over a growing wavy membrane enhances diffusive transport of solute towards and across its surface, owing to the tortuous paths of the fluid.
Such wavy membranes would be the best candidates for the origin of life, from a point of view of exchange of chemicals between the membrane and the surrounding environment.  The irregular surface would ensure that even when the external  flow is parallel to the membrane, there would be a strong transverse exchange, much faster than molecular diffusion.
Our finding of enhanced diffusion thus helps to explain the `leakiness' present in proto-cells before LUCA that they may have inherited from chemical gardens. 

Reactive interfaces play in general an important role in many chemical \cite{cardoso1,cardoso2}, biological, and engineering processes in instances ranging from heterogeneous catalysis \cite{ertl08}  to protein activity in biological membranes \cite{Bystrom2008, Sheikh2011} and to environmental flows\cite{cardoso3}. For liquid--liquid interfaces, most chemical research focuses on immiscible liquids such as certain polymer melts \cite{fredrickson1996,lyu1999} and emulsions because sharp interfaces between miscible liquids are difficult to generate and swiftly decay owing to diffusion and other transport phenomena. This limitation is eliminated or at least weakened if the two liquids produce a gel-like or solid reaction product that effectively compartmentalizes the system. Such resulting reaction products trace the macroscopic shape of the original solution interface but can reveal interesting complexity at microscopic length-scales that result from the unusually steep concentrations during their formation. Furthermore, the separating membrane or wall enforces its own rules on the evolution of the reaction systems as it controls  transmembrane transport according to reactant-specific permeabilities. Prime examples of such phenomena are a class termed chemobrionic systems, which include the so-called chemical gardens \cite{barge2015}.

Classical chemical gardens \cite{toth07,fryfogle15,kiehl15} consist of hollow precipitate tubes that form when a metal salt seed is placed into aqueous solutions of silicate, carbonate, phosphate, borates, and some other anions. The thin, hollow tubes typically consist of an outer layer of silica (in silicate systems) and an inner layer rich in metal hydroxides or oxides. Numerous chemical modifications have been reported including tube-forming reactions involving polyoxometalates \cite{cronin2012} and organic reactants \cite{pampalakis2016}.  The seed particle may be replaced by  injected salt solutions \cite{thouvenel}. Furthermore, one can reduce the dimensionality of the system by  the injection of salt solutions from a point-like source into a thin, horizontal layer of silicate solution contained within a Hele-Shaw cell \cite{haudin2014,haudin2015a,haudin2015b}.  This approach was taken further by studying the formation of precipitate membranes in a microfluidic device at the planar interface between two parallel, cocurrent reactive fluid streams \cite{batista2015}. The product wall in these experiments is analogous to the tube wall in three-dimensional chemical gardens but provides excellent control over the experimental parameters and also allows a direct view of the wall cross-section. As found in an earlier study that measured the dry weight of tubes formed over different reaction times \cite{roszol2011}, the wall thickness increases as the square root of time, thus suggesting a diffusion-controlled mechanism. In addition, the wall thickening occurs strictly in the direction of slightly acidic, metal salt solution and not towards the alkaline silicate solution. While this observation was originally made for silicate systems, the results from microfluidic growth experiments confirmed this finding for precipitation reactions between sodium hydroxide and various metal salt solutions including magnesium, manganese, iron, cobalt, and copper \cite{batista2015}.
Here we present  experimental measurements on the growth of precipitate membranes in a microfluidic system together with  an analytical model that  captures key aspects of the growth dynamics. Our results show that the small variations in the membrane thickness and the presence of a porous gel layer create a  fluctuating component of the flow velocity which, compared to pure diffusion, yields a faster growth of the precipitate.

\section{Experimental Methods}

\begin{figure}[tb]
\begin{center}
\includegraphics[width = \columnwidth]{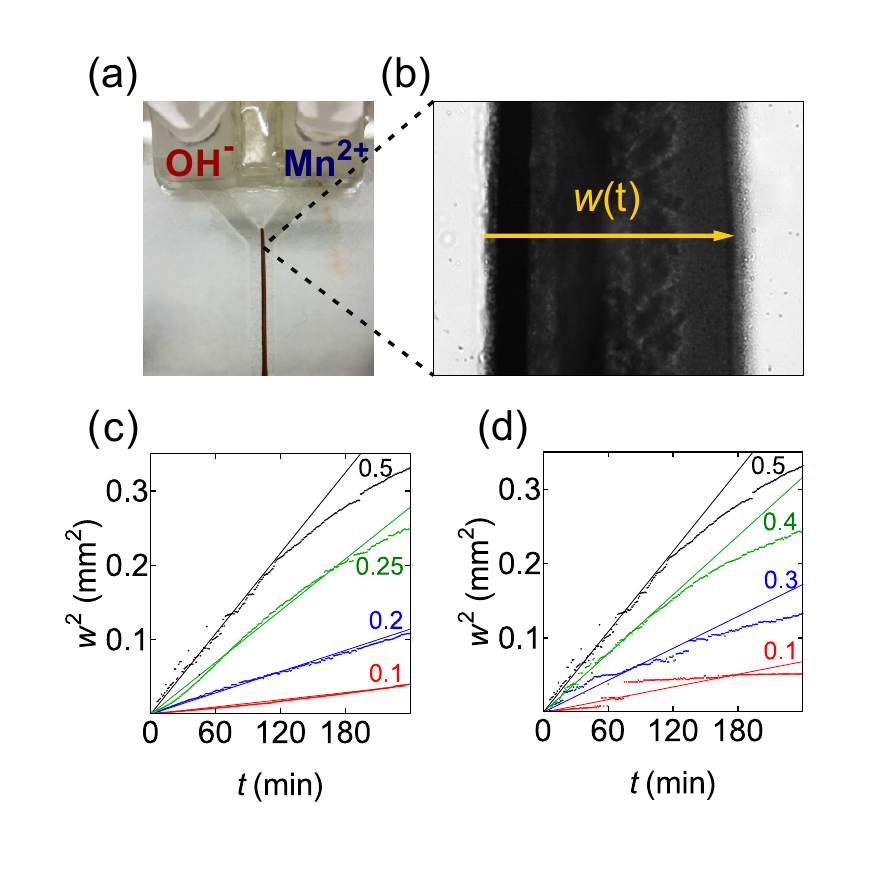}
\end{center}
\caption{(a) Y-shaped microfluidic device used for the production of a precipitate membrane (brown line) at the interface of NaOH and MnCl$_2$ solutions. (b) Optical micrograph of the membrane (black region). Its average thickness is  $w(t)$, where $t$ denotes the elapsed reaction time. (c) Evolution of the squared membrane width for four different concentrations of Mn$^{2+}$ (0.1, 0.2 0.25 and 0.5 M) and [OH$^{-}$]~= 0.5~M. (d) Evolution of the squared membrane width for four different concentrations of OH$^{-}$  (0.1, 0.3, 0.4 and 0.5 M)  and [Mn$^{2+}$]~= 0.5~M. The straight lines in (c) and (d) are linear regressions that yield concentration-dependent slopes $D_{e}$.
\label{fig:mainfacts}}
\end{figure}

Our experiments employ a Y-shaped microfluidic device that has been described in detail in reference \cite{batista2015} (Fig.~\ref{fig:mainfacts}a). It is constructed from pairs of metacrylate plates (5$\times$4~cm$^2$) sandwiching a pre-cut parafilm membrane (approximate thickness 130~$\mu$m). The cut-outs are prepared by a computer-controlled cutting tool (Silhouette Portrait) and define two inflow channels that combine to the  reaction channel. The latter is 2~cm  long and 3~mm wide, yielding a total volume of about 8~$\mu$L. The reactant solutions, NaOH and MnCl$_2$ with concentrations in the range 0--0.5~M, are delivered by a syringe pump (KD Scientific 200) through tubing and barb fittings glued to holes on the top plate. Prior to the experiment, the device channels are pumped full of water. Once the reaction solutions reach the device, the pump rate is set to a constant value of 2~mL/h per syringe. We monitor the formation of the precipitate using a charge-coupled device camera (COHU 2122) and subsequently analyze the video frames using in-house Mathematica scripts.  All experiments are carried out at room temperature.

\begin{figure}[tb]
\begin{center}
\includegraphics[width = 0.8\columnwidth]{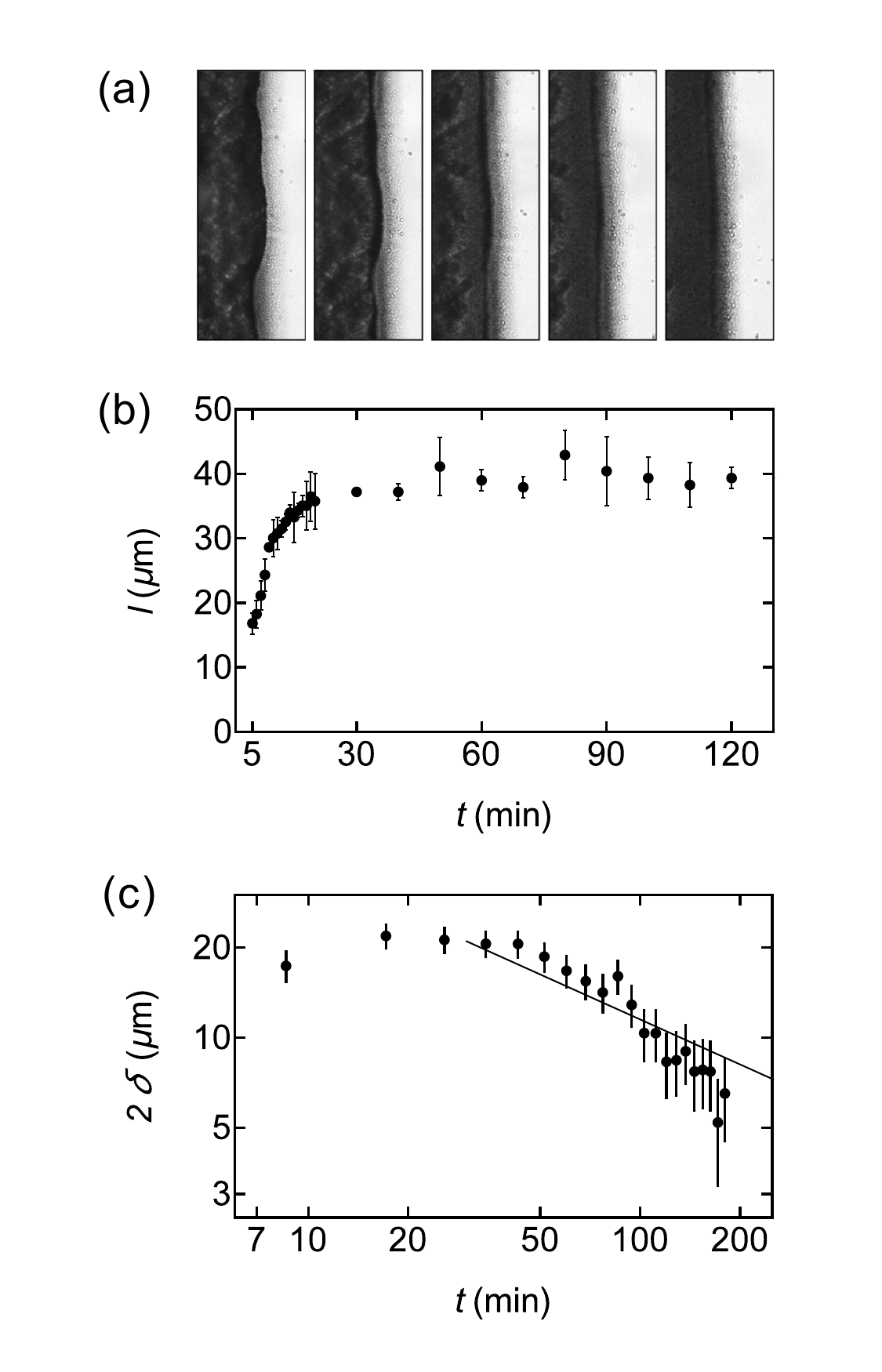}
\end{center}
\caption{(a) Micrograph sequence of the precipitate surface facing the Mn$^{2+}$ solution and the accompanying gel-like layer. Field of view: $480 \times 180$~$\mu$m$^2$. Time between frames: 25~minutes. (b) Width of the gel-like layer as a function of the elapsed reaction time. (c) Double-logarithmic plot of the amplitude of the width variations of the gel-like layer as a function of the elapsed reaction time. The linear regression line $\delta(\mu\textrm{m}) = 115/ \sqrt{t(\textrm{min})}$ is shown.
\label{fig:expresults}}
\end{figure}

\section{Experimental Results}

The parallel flow of the two solutions along the reaction channel induces the formation of a nearly linear precipitate membrane (Fig.~\ref{fig:mainfacts}a). Its width $w(t)$ is easily monitored and increases only in the direction of the Mn$^{2+}$ solution, while its interface with NaOH solution remains stationary (Fig.~\ref{fig:mainfacts}b). Figure~\ref{fig:mainfacts}c graphs the square of this width as a function of time for four different concentrations of MnCl$_2$. The data are well described by the relation $w^2 = D_{e} t$, where $D_{e}$ is an effective diffusion coefficient. Deviations from this simple dependence are observed only for high reactant concentrations and long reaction times but, even in the case of [MnCl$_2$]~= 0.5~M, the square-root dependence holds well for the first two hours of the experiment.

The surface of the membrane facing the Mn$^{2+}$ solution shows a thin layer that is distinctly different from the opaque main part of  the precipitate membrane. Based on our optical micrographs, this layer is translucent and appears somewhat granular (Fig.~\ref{fig:expresults}a). We interpret it as having a gel-like structure with large pores that allow for some fluid flow. The image sequence in Fig.~\ref{fig:expresults}a shows that the layer is present throughout all stages of the growth process. No comparable feature is observed on the hydroxide side of the precipitate membrane. The image sequence also reveals that the manganese side of the earlier membrane is not perfectly planar. Furthermore, these small variations in the membrane width decay over time, ultimately yielding very straight boundaries between the solid precipitate, its gel-like zone, and the manganese solution.

Quantitative details of these two features are illustrated in Fig.~\ref{fig:expresults}b,c. The data in Fig.~\ref{fig:expresults}b show the width of the gel-like layer as a function of the elapsed reaction time revealing an essentially constant value of $\sim 40\pm5$~$\mu$m for $t >$~30~min. During the first 30 minutes, the width increases smoothly suggesting the decay of slow transients. The double-logarithmic graph in Fig.~\ref{fig:expresults}c shows the evolution of the amplitude of the width variations as detected from images including those in Fig.~\ref{fig:expresults}a. We approximate the data set for $t >$~30~min by a power law $\delta(\mu\textrm{m}) = 115/ \sqrt{t(\textrm{min})}$, represented as the continuous line in Fig.~\ref{fig:expresults}c; this approximation lies within the error bars of experimental uncertainty and allows us to obtain an analytical solution in the modelling below.

\section{Model and Comparison with Experiments}

	\begin{figure}[tb]
		\begin{center}
			\includegraphics[width = \columnwidth]{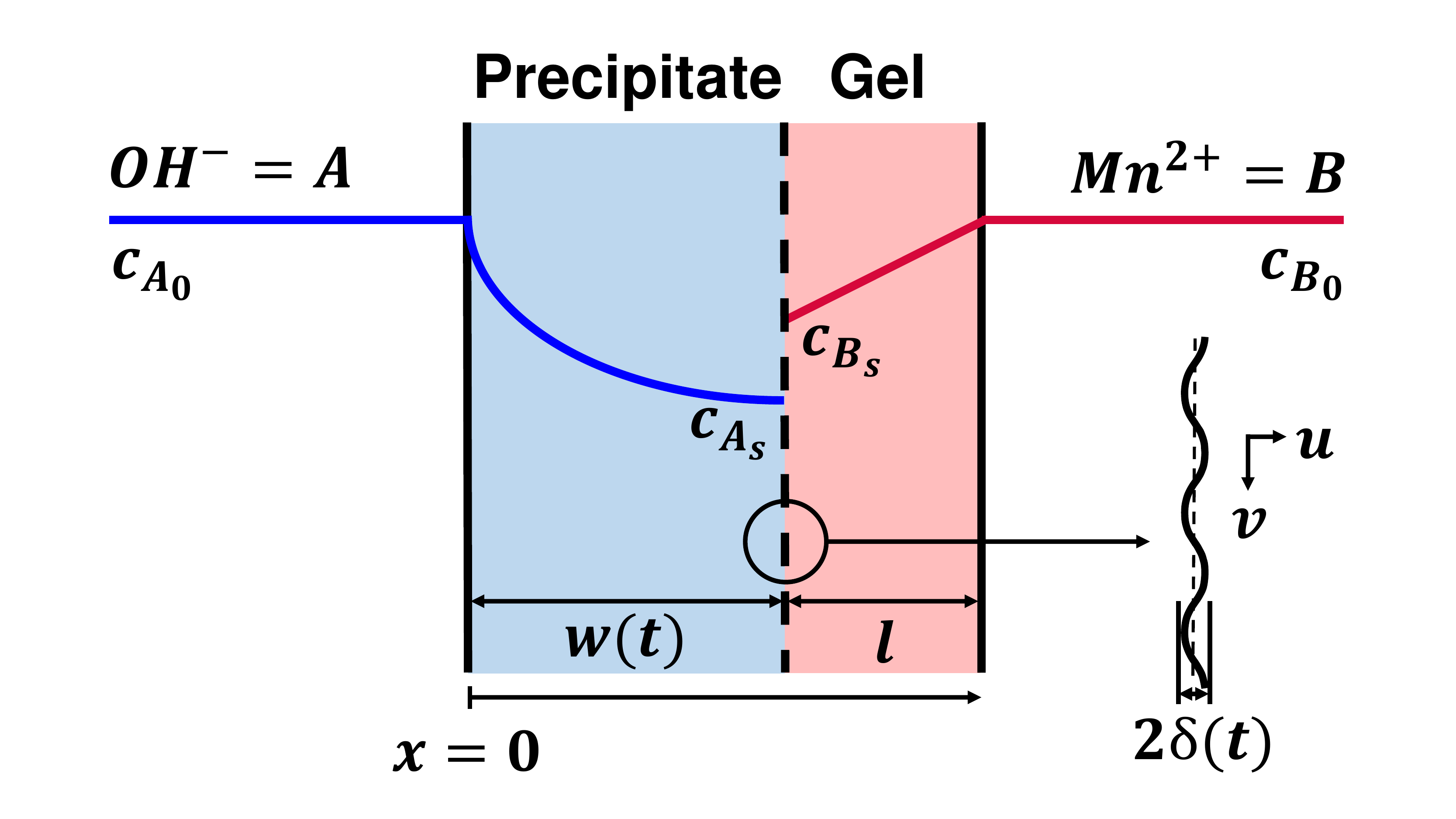}
		\end{center}
		\caption{Sketch of the concentration profiles of A and B in the precipitate and gel layers.
		\label{fig:sketch}}
	\end{figure}

				\begin{figure}[tb]
		\begin{center}
			\includegraphics[width = 0.8\columnwidth]{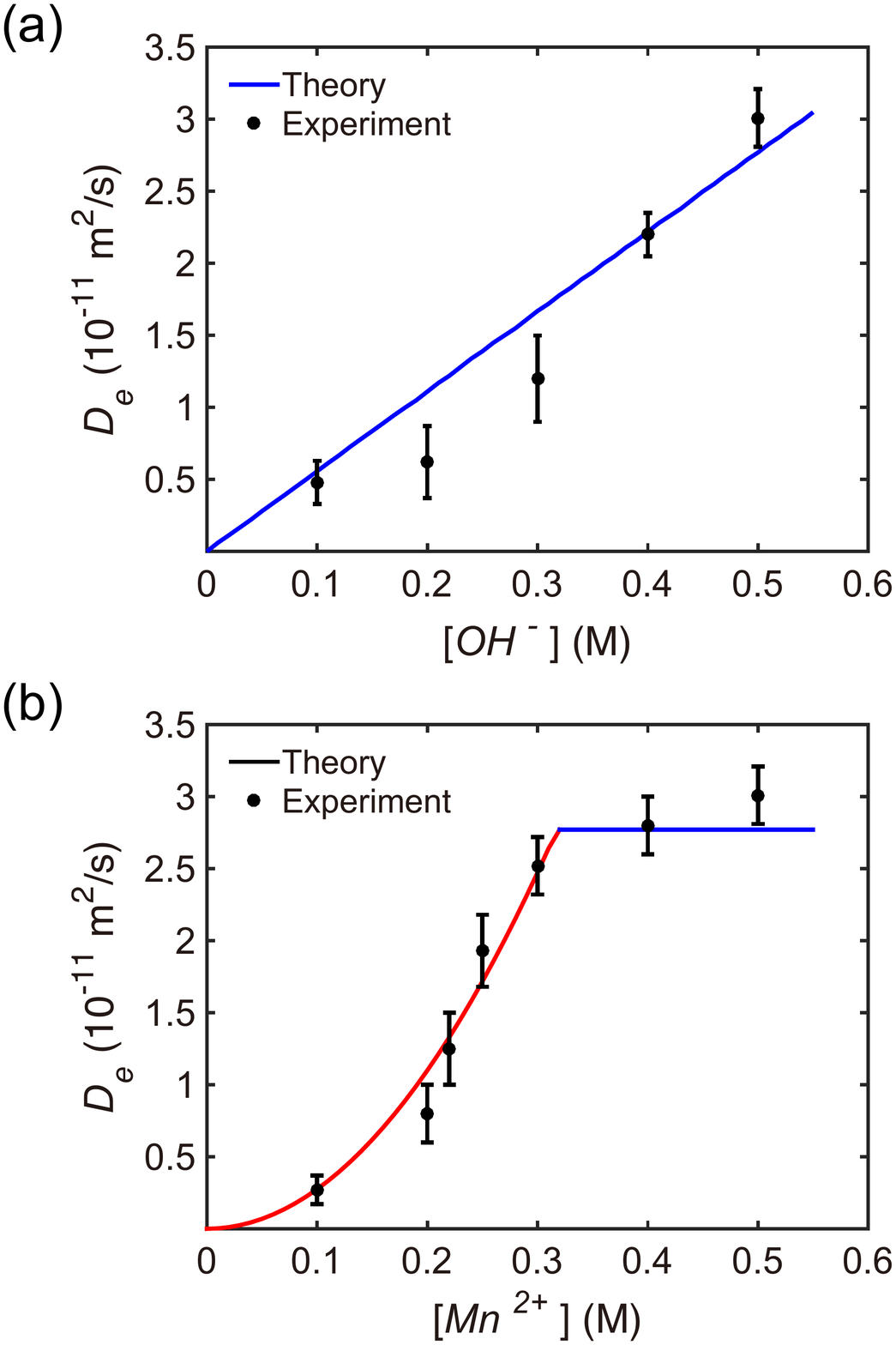}
		\end{center}
		\caption{Variation of the effective diffusivity with the concentrations of A and B. Experimental data  are shown with the prediction of Eq.~(\ref{de})
for $D_A=9.8 \times 10^{-10}$ m$^2$/s and $D_B($m$^2$/s$)=5.8 \times 10^{-7}/ \sqrt{t(\textrm{s})}$. Regimes of growth controlled by the transport of either the anion (blue) or the cation (red) to the reaction surface are shown. 
		\label{fig:model}}
	\end{figure}

Now consider the parallel flow of two aqueous solutions, one containing anion A$^{n-}$ at concentration $c_{A_0}$ and the other cation B$^{m+}$ at concentration $c_{B_0}$.  The reaction
\begin{equation}
		|\nu_{A}|\textrm{A} (aq) +|\nu_{B}|\textrm{B} (aq) \to|\nu_{C}|\textrm{C} (s)
\end{equation}
	occurs at the interface between the two streams to form a precipitate layer of product C, whose thickness $w(t)$ grows with time.  A gel layer of dissolved precipitate $C (\textrm{aq})$ develops at the solid
precipitate surface (Fig.~\ref{fig:sketch}), growing rapidly to a steady state of thickness $l\ll w(t)$. A balance of the two chemical species across the precipitate and gel layers requires that
\begin{subequations}
	\begin{eqnarray} 
		&& \frac{\partial{c_A}}{\partial{t}}=D_{A}\frac{\partial^{2}c_{A}}{\partial{x^2}},
		\label{governing-eqns1} \\
		&& \frac{\partial{c_B}}{\partial{t}}=D_{B}\frac{\partial^{2}c_{B}}{\partial{x^2}}.
		\label{governing-eqns2}
		\end{eqnarray}\label{governing-eqns}
\end{subequations}	
Here, $c_A$ and $c_B$ are the concentration of A in the pore fluid in the precipitate and the concentration of B in the gel, respectively; $D_A$ is the diffusivity of A in the porous precipitate and
$D_B$ is the diffusivity of B in the gel layer.  The transverse spatial position is denoted by $x$.
	We assume the precipitate is impermeable to the metal ion and that the reaction is almost instantaneous, so that reaction and precipitation occur only at the plane of contact of two ions. The
initial conditions at $t=0$ are that $c_A=c_{A_0}$ for $x<0$  and $c_B=c_{B_0}$ for $x>0$.  At the boundaries of the precipitate and gel layers, $x=0$ and $x=w(t)+l$, these concentrations remain constant
with time owing to the relatively fast flow in the injected streams, so that $c_A=c_{A_0}$  and $c_B=c_{B_0}$ respectively.
	The boundary conditions at the reaction plane $x=w(t)$ express conservation of total mass and the stoichiometric constraint
\begin{subequations}
	\begin{eqnarray}
		\rho_s S \frac{dw}{dt}&=&\frac{dm_A}{dt}+\frac{dm_B}{dt} \\
		&=& S \left( -D_A M_A\frac{dc_A}{dx}+D_B M_B\frac{dc_B}{dx} \right),\label{mass1} \nonumber \\
		-|\nu_B|D_A\frac{dc_A}{dx}&=&|\nu_A|D_B\frac{dc_B}{dx}.\label{mass2}
	\end{eqnarray}\label{mass}
	\end{subequations}
	Here, $S$ is the surface area of the reaction plane, $\rho_s$ is the bulk density of the solid precipitate, $m$ denotes mass in the precipitate, and $M$ the molar mass.
	Owing to the instantaneous reaction, we also impose that at least one of the concentrations at the precipitation plane, $c_{A_s}$ or $c_{B_s}$, is zero.  This condition then separates two regimes,
one in which transport is limited by the supply of A ($c_{A_s}=0$) and the other in which the growth is limited by the transport of B ($c_{B_s}=0$) to the reaction site.

		\begin{figure}[tb]
	\begin{center}
			\includegraphics[width = 0.7\columnwidth]{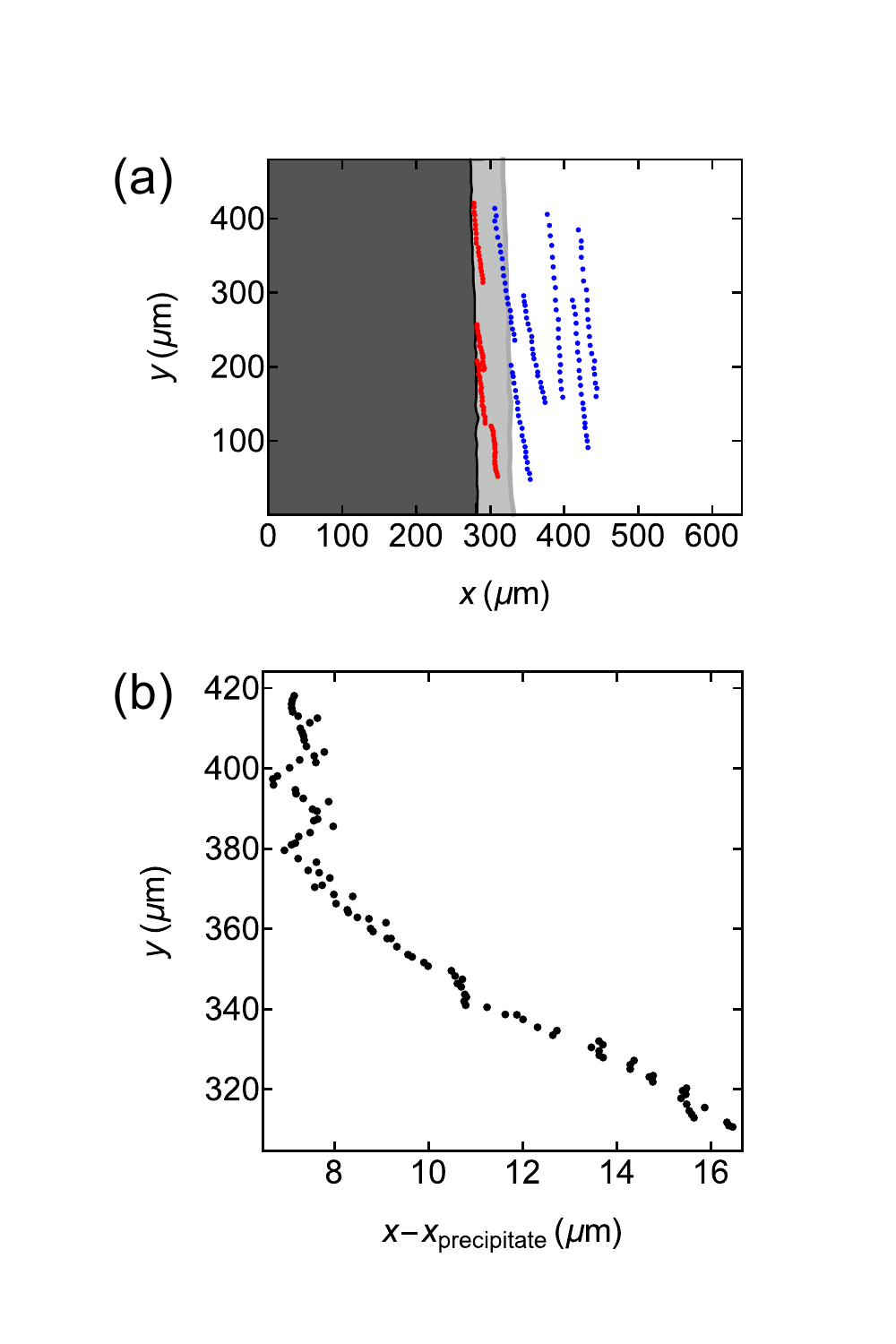}
		\end{center}
		\caption{Visualization of fluid flow near the precipitate surface.  (a) Examples of micro-bubble pathlines in the gel layer (red) and in the manganese stream (blue) during a period of time of 10~s. (b) Bubble motion in the gel layer relative to the precipitate surface. The bubble trajectory was measured after approximately 3 h of precipitate growth for [MnCl$_2$] = [NaOH] = 0.5~M.
		\label{fig:paths}}
	\end{figure}

	For sufficiently long times, such that the gel layer has attained constant thickness and the concentration profile of B therein is linear, the integration of Eqns~(\ref{governing-eqns}) with the initial and boundary conditions leads to the following
analytical relations for the effective diffusivity $D_e$ of A and B in the system and the concentration of the ion in stoichiometric excess at the reaction plane:
\begin{subequations}
	\begin{eqnarray} \label{de}
		&&\sqrt{{D_{e} \over  {4D_A}}} \cdot \exp \left({D_{e} \over {4 D_A}} \right)\cdot \textrm{erf} \sqrt{{D_{e} \over  {4D_A}}} 		\label{de1} \\
		&&=\left(M_A+M_B\frac{|\nu_B| }{|\nu_A|}\right)\frac{c_{A_0}-c_{A_s}}{\sqrt{\pi}\rho_s}, \nonumber \\
		&&D_{e}=\left[{{D_B \sqrt{t}} \over { l}} \left(M_A\frac{|\nu_A|}{|\nu_B|}+M_B \right)  \frac{2(c_{B_0}-c_{B_s})}{\rho_s} \right]^2  .  
		\label{de2}
	\end{eqnarray}
\end{subequations}

	In Fig.~\ref{fig:model} we show that almost all of the measurements for $D_e$ lie, within experimental error, on the lines predicted by Eq.~(\ref{de}) for $D_A=9.8 \times 10^{-10}$ m$^2$/s and $D_B($m$^2$/s$)=5.8 \times 10^{-7}/ \sqrt{t(\textrm{s})}$. The diffusivity of A in the precipitate was estimated from its molecular diffusivity in solution $D_{A_\infty}=5.3 \times 10^{-9}$  m$^2$/s, the measured porosity of the precipitate
$\phi=0.52$ and its tortuosity $\tau=2.8$, as $D_A=\phi D_{A_\infty}/\tau$; the relatively high value for the tortuosity is consistent with a precipitate composed of channels at an angle of $\sim 20^ \circ$ to the membrane surface \cite{koza2012}.
	For the transport of B in the gel layer, we need to consider hydrodynamic effects.  The experiments show that the precipitate has a rough surface with protrusions into the gel. The amplitude of the
protrusions $\delta$ decays  with time, from approximately 20 to 1 $\mu$m (Fig.~\ref{fig:expresults}), following the approximate dependence $\delta(\mu\textrm{m}) = 115/ \sqrt{t(\textrm{min})}$.  The stream of B is injected
with velocity $v \sim 0.0015 $ m/s and drags the fluid in the gel layer over the protrusions. Evidence of this motion is seen by tracking the paths of a few bubbles in the gel layer and in the outer stream (Fig.~\ref{fig:paths}).  The bubbles move with longitudinal speeds of approximately 8.5  and 19 $\mu$m/s in the gel and injected stream, respectively. The bubble speed in the stream is much slower than the average fluid speed because, owing to buoyancy, they lie close to the upper surface of the top plate of the micro-channel. Importantly, our observations show that the bubble paths are tortuous in the gel layer, indicating that the fluid there is moving to and fro the membrane surface (Fig.~\ref{fig:paths}b).  The bubble transverse speed in the irregular part of its path is approximately 2.4 $\mu$m/s; this suggests that the fluid is moving to and fro the membrane with a horizontal speed  $u \sim 2.4 \times 10^{-4}$ m/s.   The forced flow over the wavy topography and within the porous structure of the gel is responsible for this transverse mixing. We therefore expect transverse dispersive transport  of B of the order of $D_B \sim  a v \delta $ \cite{phillips}, where the
constant $a$ should be smaller than one. Comparison between the theory and the experiments suggests that $a \sim 0.06$, so that $D_B($m$^2$/s$)=5.8 \times 10^{-7}/ \sqrt{t(\textrm{s})}$.  Thus at a time of 200 min,
$D_B \sim 5.3 \times 10^{-9}$  m$^2$/s, i.e., the dispersive transport is then approximately an order of magnitude larger than the molecular diffusivity of the manganese cation in water,
$D_{B_\infty}=7.0 \times 10^{-10}$  m$^2$/s.  This prediction is in  agreement with all the experimental observations in  Fig.~\ref{fig:model}.

	We can identify two regimes of behaviour in Eq.~(\ref{de}), according to the limiting driving force for the precipitate growth, either the transport of A or B to the reaction plane.
	When transport of B is limiting, $D_e$ grows  quadratically with driving concentration difference across the gel layer; this behaviour is a consequence of the flow of fluid in the porous gel layer over the wavy-membrane topography, which induces the transverse localised mixing.  In contrast, when the transport of A determines the precipitation, the
diffusive transport grows linearly with the concentration difference across the precipitate layer.  The growth of the precipitate is maximized for the stoichiometric concentration ratio $c_{A_0} /c_{B_0}\sim \nu_A/ \nu_B$.
	
Our experimental method combined with theoretical analyses will also allow interesting studies of the concentration threshold for continuous membrane formation. An earlier study \cite{batista2014} reported this value as 0.1 mol/L for the case of iron sulfide tubes. For lower reactant concentrations, only rising plumes of individual colloidal particles were observed. This threshold phenomenon should also exist in our experimental setting which is ideally suited for systematic experiments due to the ease of observation and the system's simple geometry. Other targets for future studies include the investigation of shear rates on the wall growth and the associated waviness. There are also potential technological applications associated with enhanced diffusion across wavy membranes. One can envisage both fluid flow control via surface topography, and the opposite: pre-designed flow perturbations could be used to shape the membrane. We believe that the work presented here will provide valuable starting points for such efforts.

More broadly, we have shown that a wavy membrane and neighbouring gel layer is the essential factor driving enhanced diffusion. Both laboratory chemical gardens and hydrothermal vents in the oceans are examples of chemobrionic systems that possess such membranes. In the case of hydrothermal vents this attribute takes on a great significance because the diffusion of material across such a membrane must have been fundamental to the origin of life. Indeed, there is an apparent paradox at the heart of the earliest proto-biochemistry. A membrane had both to protect the complex chemistry from its environment yet at the same time to allow the passage of materials; to be at once permeable to some ions and so allow mixing, but at the same time to maintain steep pH gradients \cite{lane2012, sousa2013, sojo2016}. Waviness provides a plausible mechanism to achieve enhanced transport across relatively thick abiotic hydrothermal vent membranes at the very beginnings of the climb in complexity towards life. Prebiotic chemistry needed to sustain disequilibrium with its surroundings to possess pH and reduction potential by maintaining proton and redox gradients \cite{sousa2013}. The evolution of ion-tight biological membranes was contingent on the prior invention of active ion pumping, or chemiosmosis \cite{mitchell1961, sojo2016, strbak2016}. Long before the emergence of chemiosmosis with its active mechanism of ion transport, waviness constitutes a natural passive mechanism for the enhancement of transport across a semipermeable membrane that in our view forms part of the answer to the puzzle of the origin of membrane bioenergetics \cite{lane2012} and thereby the passage from passive osmosis in a hydrothermal vent vesicle to active chemiosmosis in a protocell. There is a vast distance between a chemical garden membrane and LUCA, but the fact that LUCA still needed a leaky membrane even at that much later stage in that climb in complexity \cite{sojo2016} demonstrates what a vital aspect this ion transport mechanism constituted for proto-life. `Prebiotic chemistry has been approached from the intellectual tradition of synthetic chemistry, and the apotheosis is the ``one-pot synthesis''. But cells are not simply a pot of chemicals; they have a structure in space' \cite{sojo2016}; waviness is how hydrothermal vent pores -- microfluidic reactor vessels that are marvels of natural chemical engineering -- optimize their fluid mechanics \cite{barge2016}.

	
	\begin{acknowledgments}
	SSSC acknowledges the financial support of the UK Leverhulme Trust project RPG-2015-002. JHEC acknowledges the financial support of the Spanish MINCINN project FIS2013-48444-C2-2-P. OS acknowledges support by the US National Science Foundation under Grant 1609495. BCB thanks the Brazilian National Council for Scientific and Technological Development, CNPq, for a postdoctoral fellowship.
	\end{acknowledgments}
	
\bibliographystyle{unsrt}
\bibliography{growth1}

\end{article}
	
\end{document}